\documentclass[12pt,english]{article}
\usepackage[T1]{fontenc}
\usepackage[latin1]{inputenc}
\usepackage{a4wide}
\usepackage{amsmath}
\usepackage{graphicx}
\usepackage{amssymb}

\makeatletter
\usepackage{graphicx}
\usepackage[T1]{fontenc}
\usepackage[latin1]{inputenc}
\makeatletter

\usepackage{babel}

\usepackage{babel}
\makeatother
\begin{document}
\begin{flushright}APCTP-2003-003\end{flushright}

\begin{flushright}December 2003\end{flushright}

\vspace{2cm}

\begin{center}{\LARGE Excited states NLIE for sine-Gordon model}\end{center}{\LARGE \par}

\begin{center}{\LARGE in a strip with Dirichlet boundary conditions}\end{center}{\LARGE \par}

\vspace{1.5cm}

\begin{center}{\large Changrim Ahn$^{1}$, Marco Bellacosa$^{2,3}$
and Francesco Ravanini$^{2}$}\end{center}{\large \par}

\vspace{0.7cm}

\begin{center}\textit{$^{1}$Ewha Womans University, Seoul 120-750,
Korea}\end{center}

\begin{center}\textit{$^{2}$INFN, Sez. di Bologna, Italy}\end{center}

\begin{center}\textit{$^{3}$Dottorato di Ricerca in Fisica, Bologna
University, Italy}\end{center}

\vspace{1cm}

\begin{abstract}
We investigate various excited states of Sine-Gordon model on a strip
with Dirichlet boundary conditions on both boundaries using a Non
Linear Integral Equation (NLIE) approach.

\newpage
\end{abstract}

\section{Introduction}

Consider the quantum field theory of a self-interacting scalar boson
$\phi(x,t)$ on a 1+1 dimensional strip, infinite in time direction
and finite in space with bulk action \begin{equation}
\mathcal{A}_{DSG}={\displaystyle \frac{1}{2}\int_{-\infty}^{+\infty}{\rm d}t{\displaystyle \int_{0}^{L}\textrm{d}x\left[\left(\partial_{t}\phi\right)^{2}-\left(\partial_{x}\phi\right)^{2}+{\displaystyle \frac{m_{0}^{2}}{\beta^{2}}\cos\beta\phi}\right]}}\label{eq:2}\end{equation}
 and with the Dirichlet boundary conditions $\phi\left(0,t\right)\equiv\phi_{-}+\frac{2\pi}{\beta}m_{-}$
and $\phi(L,t)=\phi_{+}+\frac{2\pi}{\beta}m_{+}$, $m_{\pm}\in\mathbb{Z}$.
We shall refer to this integrable theory as the Dirichlet sine-Gordon
(DSG) model. It has several important applications ranging from condensed
matter physics to string theory.

The well known bulk particle spectrum of sine-Gordon, composed of
solitons and antisolitons with topological charge 1 and $-1$ respectively,
and, only in the \emph{attractive} regime $0<\beta\leq\sqrt{4\pi}$,
bound states of solitons known as breathers, is complemented by the
rich structure of boundary bound states described in \cite{dorey-mattson}.
The expressions for the bulk S-matrices can be found in \cite{zam-zam},
those for the fundamental soliton or antisoliton reflection matrices
are given in \cite{ghoshal-zam} and the general boundary bound state
excited reflection matrices can be found in \cite{dorey-mattson}.

An important feature of DSG is the conservation of the topological
charge\begin{equation}
Q\equiv\frac{\beta}{2\pi}\left[\int_{0}^{L}dx\frac{\partial}{\partial x}\phi(x,t)-\phi_{+}+\phi_{-}\right]=m_{+}-m_{-}\in\mathbb{Z}\quad.\label{eq:Q}\end{equation}
The model enjoies the discrete symmetry of the field $\phi\to\phi+\frac{2\pi}{\beta}m$
and simultaneously $\phi_{\pm}\to\phi_{\pm}+\frac{2\pi}{\beta}m$
($m\in\mathbb{Z}$). The charge conjugation symmetry $\phi\to-\phi$
sending solitons into antisolitons is also guaranteed provided $\phi_{\pm}\to-\phi_{\pm}$
simultaneously. It sends $Q\to-Q$, so one can restrict attention
to the study of positive $Q$ and then act with this transformation
to obtain states of negative $Q$. The periodicity allows to restrict
the boundary parameters to the range $0\leq\phi_{\pm}<\frac{2\pi}{\beta}$.%
\footnote{Notice that, unlike in the single boundary case of \cite{ghoshal-zam,saleur-skorik,dorey-mattson},
in the present two boundaries model we cannot further restrict both
boundary parameters to $0\leq\phi_{\pm}<\frac{\pi}{\beta}$.%
}

A problem of great interest is to connect the scattering theory approach
just mentioned to the somewhat complementary description of perturbed
conformal field theory data. In this letter we attack this problem
from the point of view of the nonlinear integral equation (NLIE) approach
which was developed for vacuum scaling functions in \cite{klumper,ddv92,ddv95}
and extended later to the excited states \cite{mariottini,ddv97,FRT1,FRT2,FRT3}.
In this framework, exact scaling functions of finite size effects
provide a way to investigate the flows from ultraviolet (UV) to infrared
(IR) scales in integrable quantum field theory, hence building a bridge
between the perturbed CFT description and the factorized scattering
one. The NLIE for the vacuum Casimir energy was already deduced, along
a similar line to those we shall illustrate here, some years ago in
\cite{LSSM}. Our interest here is to establish a general NLIE valid
for all states in the finite size spectrum. This allows to identify
{}``particle'' states in the scattering description with {}``conformal''
states in the perturbed CFT description. 

To make this approach viable, one should start from an exact solution
of a lattice regularization of the model. This is normally provided,
in NLIE approach, by some 2D light-cone vertex model or equivalently
by some inhomogeneous 1D spin chain. By suitable scaling limit the
continuum renormalized theory can be reached. The Bethe ansatz equations
that solve the lattice model are turned in the continuum limit into
a NLIE that has to be considered as the basic renormalized tool to
calculate the eigenvalues of the integral of motions of the theory
on any state in finite volume.

The homogeneous antiferromagnetic XXZ spin 1/2 model in a chain of
$N$ sites with lattice spacing $a$, coupled to parallel magnetic
fields $h_{+}$ and $h_{-}$ at the left and right boundaries respectively,
has Hamiltonian

\begin{equation}
\mathcal{H}(\gamma,h_{+},h_{-})=-J\sum_{n=1}^{N-1}\left(\sigma_{n}^{x}\sigma_{n+1}^{x}+\sigma_{n}^{y}\sigma_{n+1}^{y}+\cos\gamma\sigma_{n}^{z}\sigma_{n+1}^{z}\right)+h_{-}\sigma_{1}^{z}+h_{+}\sigma_{N}^{z}\quad.\label{eq:1}\end{equation}
 Here $\sigma_{n}^{\alpha}$, $\alpha=x,y,z$ are Pauli matrices and
$0\leq\gamma<\pi$. Where convenient, we shall also use the equivalent
parameter $p=\frac{\pi}{\gamma}-1$, whose range is $0<p<\infty$.
The Hamiltonian (\ref{eq:1}), as well as its inhomogeneous generalizations,
can be constructed in a double row transfer matrix framework. For
details see \cite{sklyanin}. 

The bare continuum limit $N\rightarrow\infty$, $a\rightarrow0$ while
$Na=L$ remains fixed, is known to give, in the homogeneous case (\ref{eq:1})
a massless free boson $\phi(x)$ compactified on a circle of radius
$R=1/\sqrt{2(\pi-\gamma)}$ \cite{luther-peschel}. Among the many
possible deformations of the Hamiltonian (\ref{eq:1}) leading to
sine-Gordon in the bare continuum limit, we choose the one introducing
alternating inhomogeneities $\vartheta_{n}=(-1)^{n}\Theta$ in the
sites of the chain. This choice has been known for sometime \cite{ddv87,saleur-resh}
to give a correct construction of sine-Gordon model in the bulk in
cylindrical geometry, when the appropriate scaling limit is chosen,
with periodic or twisted boundary conditions. It is then natural to
expect that the same construction in the presence of boundary magnetic
fields $h_{\pm}$ can also provide an effective tool to define the
renormalized DSG theory. It is worthwhile to recall that, as the homogeneous
XXZ chain is equivalent to a 6-vertex model on a square lattice, this
modified XXZ chain is also equivalent to a 6-vertex model, but --
as a consequence of the introduced inhomogeneities -- defined on a
lattice rotated by 45°, i.e. on what can be thought as a Minkovski
space discretized along the light-cone directions. This is why this
construction is often referred as \emph{light cone lattice} construction
of the sine-Gordon model \cite{ddv87}. For $\Theta\to\infty$, $N\to\infty$
and $a\to0$ while $L=Na$ is fixed, contact can be made, along lines
similar to those in \cite{destri-segalini}, with the lagrangean formulation
of DSG model, eq.(\ref{eq:2}). The XXZ anisotropy $\gamma$ is related
to the SG coupling $\beta$ by $\beta^{2}=8(\pi-\gamma)=\frac{8\pi p}{p+1}$.
Details of this bare continuum limit are out of the scope of the present
paper.

In this preliminary letter we put our accent on the presentation of
the NLIE got from this construction and analyze only a few bulk excited
states with nonexcited boundaries, in order to show consistency with
expected results. The careful treatment of the full situation with
boundary bound states would involve more delicate issues of analytic
continuation in the boundary parameters $\phi_{\pm}$ that we choose
to postpone to a more extensive forthcoming publication \cite{preparation}
for the reasons explained at the end of sect. 2.

\section{Bethe Ansatz and NLIE}

The Bethe Ansatz equations for the boundary XXZ chain (\ref{eq:1})
have been written by Alcaraz et al. \cite{alcaraz} and Sklyanin \cite{sklyanin}
some years ago, using an algebraic approach. It is straightforward
to generalize them with the introduction of the alternating inhomogeneities
\cite{saleur-skorik}.

Eigenvalues of the double row transfer matrix can be constructed in
terms of sets of distinct numbers $\vartheta_{1},...,\vartheta_{M}$
called \emph{roots.} They are in number of $M$, ($M\leq N$) and
must satisfy the Bethe ansatz equations \begin{equation}
\left[s_{1/2}(\vartheta_{j}+\Theta)s_{1/2}(\vartheta_{j}-\Theta)\right]^{N}s_{H_{+}/2}(\vartheta_{j})s_{H_{-}/2}(\vartheta_{j})={\displaystyle \prod_{k=1,k\not=j}^{M}s_{1}(\vartheta_{j}-\vartheta_{k})s_{1}(\vartheta_{j}+\vartheta_{k})}\label{eq:3}\end{equation}
 where\[
s_{\nu}(x)=\frac{\sinh\frac{\gamma}{\pi}(x+i\nu\pi)}{\sinh\frac{\gamma}{\pi}(x-i\nu\pi)}\]
 and $H_{\pm}$ is defined such that $h_{\pm}=\sin\gamma\cot\frac{\gamma}{2}(H_{\pm}+1)$
and we choose as fundamental region $-p-1<H_{\pm}<p+1$. Notice that
the boundary terms in the Bethe equations disappear when $H_{\pm}=0$,
i.e. $h_{\pm}=1+\cos\gamma\equiv h_{c}$. In such case, as it was
shown in \cite{pasquier-saleur,ddv-sos}, the system becomes $SL_{q}(2)$
invariant. 

The antiferromagnetic vacuum turns out to be a maximal set $M=\frac{N}{2}$
of real roots and it exists for $N$ even only. In the region $0\leq\gamma<\pi$
of interest for us, and for small enough boundary magnetic fields,
this is the true ground state of the theory. For $N$ odd instead
the states with lowest possible total spin have $M=\frac{N-1}{2}$
roots and one hole. However, to deal correctly with the continuum
limit one has to consider \emph{both} $N$ even and odd sectors, like
it was shown in the periodic case in \cite{FRT1,FRT2,FRT3}. The symmetry
of (\ref{eq:3}) $\{\vartheta_{j}\}\rightarrow\{-\vartheta_{j}\}$,
evident from the Bethe equations, implies that only roots with positive
real part are independent parameters characterizing a Bethe state.
The value $\vartheta_{j}=0$ is a solution of (\ref{eq:3}) for any
$N$ and $M$. However, the corresponding Bethe state would vanish,
so one has always to subtract this unwanted root, i.e. to create a
hole at $\vartheta=0$.

The domain of root distribution can be considered as a semistrip $\mathbb{U}_{+}$
of the complex $\vartheta$ plane:\[
\mathbb{U}_{+}=\left\{ \vartheta\in\mathbb{C}\,|\,\mathrm{Re}\vartheta>0\,,\,-\frac{\pi^{2}}{2\gamma}<\mathrm{Im}\vartheta\leq\frac{\pi^{2}}{2\gamma}\,\mathrm{or}\,\mathrm{Re}\vartheta=0\,,\,0<\mathrm{Im}\vartheta<\frac{\pi^{2}}{2\gamma}\right\} \quad.\]
 This also excludes another unwanted root at $i\frac{\pi^{2}}{2\gamma}$
and considers only half of the imaginary axis, as it should for symmetry.
However, for computational purposes, it is often better to double
this strip by mirroring all the roots \[
\mathbb{U}=\left\{ \vartheta\in\mathbb{C}\,|\,\mathrm{Re}\vartheta\in\mathbb{R}\,,\,-\frac{\pi^{2}}{2\gamma}<\mathrm{Im}\vartheta\leq\frac{\pi^{2}}{2\gamma}\right\} \quad.\]
 To each root $\vartheta_{j}$ associate its mirror root $\vartheta_{-j}\equiv-\vartheta_{j}$.
Define the function

\begin{equation}
\varphi_{\nu}\left(\vartheta\right)\equiv\pi+i\log s_{\nu}(\vartheta)\label{eq:4}\end{equation}
 with the oddity condition $\varphi_{\nu}(-u)=-\varphi_{\nu}(u)$
fixing the fundamental branch of the logarithm. It is periodic in
the imaginary direction, with period $i\frac{\pi^{2}}{\gamma}$, and
real on the real axis. We choose as fundamental periodicity the strip
$\vartheta\in\mathbb{U}$. Singularities of this function are distributed
along the imaginary axis:\[
\mathrm{Re}\vartheta=0,\qquad\mathrm{Im}\vartheta=\pm\pi(k(p+1)-\nu),\quad k\in\mathbb{Z}\]
 so that the fundamental analyticity strip is limited to $|\mathrm{Im}\vartheta|<\pi\min(\nu,p+1-\nu)$.

In terms of the function (\ref{eq:4}) the logarithm of the Bethe
equations (\ref{eq:3}) can be expressed as\[
\begin{array}{c}
N\left[\varphi_{1/2}(\vartheta_{j}+\Theta)+\varphi_{1/2}(\vartheta_{j}-\Theta)\right]+\varphi_{H_{+}/2}(\vartheta_{j})+\varphi_{H_{-}/2}(\vartheta_{j})\\
{\displaystyle +\varphi_{1}(\vartheta_{j})+\varphi_{1}(2\vartheta_{j})-\sum_{k=1}^{2M}\varphi_{1}(\vartheta_{j}-\vartheta_{k})=2\pi I_{j}}\quad,\quad I_{j}\in\mathbb{Z}\quad.\end{array}\]
 Eigenvalues of the transfer matrix can be expressed in terms of its
roots. In the following, we shall be mainly interested in the energy
spectrum, for which the formula is\begin{equation}
E=-{\displaystyle \frac{1}{2a}{\displaystyle \sum_{k=1}^{M}\left({\displaystyle \frac{\textrm{d}}{\textrm{d}\Theta}\varphi_{1/2}\left(\Theta-\vartheta_{k}\right)+{\displaystyle \frac{\textrm{d}}{\textrm{d}\Theta}\varphi_{1/2}\left(\Theta+\vartheta_{k}\right)}}\right)}}+\frac{1}{a}\frac{\textrm{d}}{\textrm{d}\Theta}\varphi_{1/2}\left(\Theta\right)\quad.\label{eq:24}\end{equation}

Define for $\vartheta\in\mathbb{U}$ the so called \emph{counting
function}

\begin{align}
Z_{N}(\vartheta) & =N\left[\varphi_{1/2}(\vartheta+\Theta)+\varphi_{1/2}(\vartheta-\Theta)\right]+\varphi_{H_{+}/2}(\vartheta)+\varphi_{H_{-}/2}(\vartheta)\nonumber \\
 & -{\displaystyle \sum_{k=-M}^{M}\varphi_{1}(\vartheta-\vartheta_{k})+\varphi_{1}(\vartheta)+\varphi_{1}(2\vartheta)}\label{eq:5}\end{align}
 in terms of which the logarithm of the Bethe equations simply becomes
the condition

\begin{equation}
Z_{N}\left(\vartheta_{j}\right)=2\pi I_{j}\quad.\label{eq:6}\end{equation}
 The last term in (\ref{eq:5}) takes care of the fact that in the
second member of (\ref{eq:3}) the product does not include factors
with $k=j$. The last but one instead explicitly subtracts the unwanted
root $\vartheta_{0}=0$. The integers $I_{j}$ play the role of quantum
numbers.

The analytic structure of $Z_{N}(\vartheta)$ makes it convenient
to classify the roots and related objects of the Bethe ansatz (\ref{eq:3})
as follows

\begin{enumerate}
\item \emph{real roots} $\vartheta_{k},\, k=1,...,M_{R}\,,\,\vartheta_{k}>0$:
they are strictly positive real solutions of (\ref{eq:3}) and (\ref{eq:6}); 
\item \emph{holes} $\vartheta_{k},\, k=1,...,N_{H}\,,\,\vartheta_{k}>0$:
strictly positive real solutions of (\ref{eq:6}) that are not solutions
of the Bethe Ansatz (\ref{eq:3}) ; 
\item \emph{close roots} $\vartheta_{k},\, k=1,...,M_{C}$: complex solutions
with $\mathrm{Re}\vartheta_{k}\geq0$ and imaginary part in the range
$0<|\mathrm{Im}\vartheta_{k}|<\textrm{min}\pi\left(1,p\right)$; 
\item \emph{wide roots} $\vartheta_{k},\, k=1,...,M_{W}\,,\,\mathrm{Re}\vartheta_{k}\geq0$:
complex conjugate solutions with imaginary part $\pi\textrm{min}\left(1,p\right)<|\mathrm{Im}w_{k}|<\frac{\pi^{2}}{2\gamma}$. 
\end{enumerate}
For further convenience it is useful to introduce the notion of \emph{self-conjugate}
roots, i.e. wide roots with $\mathrm{Im}\vartheta_{k}=\frac{\pi^{2}}{2\gamma}$,
whose complex conjugate is the root itself, due to the periodicity
of Bethe equations. Their number will be indicated as $M_{SC}$. Also
we shall refer to roots or holes lying on the imaginary axis as \emph{magnetic.}

The function $Z_{N}(\vartheta)$ for $\vartheta\in\mathbb{R}$ is
globally monotonically increasing. However, there may be points where
locally $\dot{Z}_{N}(\vartheta)<0$. In particular holes or roots
$s_{j}$ such that $\dot{Z}_{N}(s_{j})<0$ are called \emph{special}
holes or roots\emph{.} If a special object appears, then there must
be also two other objects (real roots or holes) with the same quantum
number, as imposed by the global increasing monotonicity of the counting
function. Moreover, as two roots with the same quantum number are
not allowed in Bethe Ansatz, at maximum one object of this triple
can be a root, the others are forced to be holes. In the following
we indicate the number of specials with $N_{S}$. A special object
should be counted both as special (i.e. in $N_{S}$) and as root or
hole (i.e. in $M_{R}$ or $N_{H}$) according to its nature.

One may relate the numbers of various types of roots to the 3rd component
of total spin of the system $S=\frac{N}{2}-M$. To do that, we express
the asymptotics of the function $Z_{N}(\vartheta)$ on the real axis
of $\vartheta$ where they can be compared with the counting of real
roots and holes. As a result we get the following \emph{counting equation}
to be satisfied by any allowed root configuration

\begin{equation}
N_{H}-2N_{S}=2S+M_{C}+2{\rm step}(p-1)M_{W}+\mathrm{step}(p-1)+\left\lfloor -\frac{2S}{p+1}-\frac{H}{p+1}\right\rfloor \label{eq:9}\end{equation}
 where $\left\lfloor x\right\rfloor $ denotes the integer part of
$x$ and $H=\frac{H_{+}+H_{-}}{2}$.

Following the standard derivation as illustrated in \cite{ddv97,FRT2,feverati},
we obtain a NLIE for the function $Z_{N}(\vartheta)$. The continuum
limit can be taken by sending $N\to\infty$ and $a\to0$ in such a
way that $L=Na$ remains constant. The only way to get a sensible
NLIE satisfied by the limiting counting function $Z(\vartheta)\equiv\lim_{N\rightarrow\infty}Z_{N}(\vartheta)$
is to admit that also $\Theta$ rescales as

\[
\Theta\sim\log\frac{2N}{\mathcal{M}L}\]
 where $\mathcal{M}$ is a mass scale. We often use in the following
the dimensionless scale parameter $l=\mathcal{M}L$. As a result of
this limit procedure, one can define the NLIE on the continuum\[
\begin{array}{ll}
Z(\vartheta) & =2l\sinh\vartheta+g(\vartheta|\{\vartheta_{k}\})+P\left(\vartheta|H_{+},H_{-}\right)\\
 & -2i\mathrm{Im}\int\textrm{d}xG(\vartheta-x-i\varepsilon)\log\left[1-(-1)^{M_{SC}}e^{iZ(x+i\varepsilon)}\right]\end{array}\]
where the boundary contribution is given by\[
P\left(\vartheta|H_{+},H_{-}\right)=2\pi\int_{0}^{\vartheta}\textrm{d}x[F(x,H_{+})+F(x,H_{-})+G(x)+J(x)]\]
 with

\begin{equation}
G(\vartheta)=\int_{-\infty}^{+\infty}\frac{\textrm{d}k}{2\pi}e^{ik\vartheta}\frac{\sinh\frac{\pi}{2}(p-1)k}{2\sinh\frac{\pi}{2}pk\cosh\frac{\pi}{2}k}\qquad\mathrm{for}\qquad|\mathrm{Im\vartheta|}<\pi\min(1,p)\label{eq:G}\end{equation}
 \[
J(\vartheta)=\int_{-\infty}^{+\infty}\frac{\textrm{d}k}{2\pi}e^{ik\vartheta}\frac{\sinh\frac{\pi}{4}(p-1)k\cosh\frac{\pi}{4}(p+1)k}{\sinh\frac{\pi}{2}pk\cosh\frac{\pi}{2}k}\qquad\mathrm{for}\qquad|\mathrm{Im\vartheta|}<\frac{\pi}{2}\min(1,p)\]
\begin{equation}
F(\vartheta,H)=\int_{-\infty}^{+\infty}\frac{\textrm{d}k}{2\pi}e^{ik\vartheta}\mathrm{sign}(H)\frac{\sinh\frac{\pi}{2}(p+1-|H|)k}{\sinh\frac{\pi}{2}pk\cosh\frac{\pi}{2}k}\qquad\mathrm{for}\qquad|\mathrm{Im\vartheta|<\frac{\pi}{2}|}H|\label{eq:16}\end{equation}
The source term is given by\[
g(\vartheta|\{\vartheta_{k}\})\equiv\sum_{k}c_{k}[\chi_{(k)}(\vartheta-\vartheta_{k})+\chi_{(k)}(\vartheta+\vartheta_{k})]\]
where

\begin{equation}
\chi(\vartheta)=2\pi\int_{0}^{\vartheta}\textrm{d}xG(x)\label{eq:chi}\end{equation}
and $\{\vartheta_{k}\}$ is the set of position of the various objects
(holes, close and wide roots, specials) characterizing a certain state.
They are characterized by the quantization rule\[
Z_{N}(\vartheta_{j})=2\pi I_{j}\quad,\quad I_{j}\in\mathbb{Z}+\frac{\rho}{2}\qquad\rho=M_{SC}\,\bmod\,2\quad.\]
The coefficients $c_{k}$ are given by\[
c_{k}=\left\{ \begin{array}{ll}
+1 & \mathrm{for\, holes}\\
-1 & \mathrm{for\, all\, other\, objects}\end{array}\right.\]
and for any function $f(\vartheta)$ we define\[
f_{(k)}(\vartheta)=\left\{ \begin{array}{ll}
f_{II}(\vartheta) & \mathrm{for\, wide\, roots}\\
f(\vartheta+i\varepsilon)+f(\vartheta-i\varepsilon) & \mathrm{for\, specials}\\
f(\vartheta) & \mathrm{for\, all\, other\, objects}\end{array}\right.\]
where the \emph{second determination} of $f(\vartheta)$ is defined
as\[
f_{II}(\vartheta)=\left\{ \begin{array}{ccc}
f(\vartheta)+f(\vartheta-i\pi\mathrm{signIm}\vartheta) & \mathrm{if} & p>1\\
f(\vartheta)-f(\vartheta-i\pi p\mathrm{signIm}\vartheta) & \mathrm{if} & p<1\end{array}\right.\,\,{\rm for}\,|{\rm Im}\vartheta|>\pi\min(1,p)\quad.\]
The contribution of special objects comes from the fact that the logarithmic
term inside the integral can go off the fundamental branch right when
$\dot{Z}_{N}<0$. In this case, the contribution of the jump in the
logarithm amounts exactly to the source term of a special object.
For large $l$ where the driving term dominates, monotonicity excludes
the presence of special holes or roots. Therefore they should not
be regarded as objects that can be added at will, but better as artefacts
that appear only for relatively small values of $l$, dictated by
the breakdown of analyticity of the equation at certain points. It
is also clear from analysis done in \cite{ddv97,FRT2,FRT4,feverati}
that the number $N_{H}-2N_{s}$ remains constant along a flow in $l$,
and equals $N_{H}$ for $l$ sufficiently large. 

For the vacuum state containing real roots only, this equation coincides
with the one found some years ago in \cite{LSSM}. Once the equation
is solved for $Z(\vartheta+i\varepsilon)$ one can use this result
to compute the $Z(\vartheta)$ function at any value in the analyticity
strip $|\mathrm{Im}\vartheta|<\pi\min(1,p)$, provided the function
$P(\vartheta|H_{+},H_{-})$ is well defined there (see comments below).
To extend the function outside this analyticity strip one has to resort
to the following modification of the NLIE

\[
\begin{array}{ll}
Z(\vartheta) & =2l\sinh_{II}\vartheta+g_{II}(\vartheta|\{\vartheta_{k}\})+P_{II}\left(\vartheta|H_{+},H_{-}\right)\\
 & -2i\mathrm{Im}\int\textrm{d}xG_{II}(\vartheta-x-i\varepsilon)\log\left[1-(-1)^{M_{SC}}e^{iZ(x+i\varepsilon)}\right]\quad.\end{array}\]

Once $Z(\vartheta)$ is known, it can be used to compute the energy.
It is composed of bulk and boundary terms whose expression can be
found in \cite{LSSM} and a Casimir energy scaling function given
by

\[
E=\mathcal{M}\sum_{k}c_{k}\cosh_{(k)}\vartheta_{k}-\mathcal{M}\int\frac{\textrm{d}x}{2\pi}\sinh xQ(x)\quad.\]

In the IR limit $l=\mathcal{M}L\to\infty$ one can make contact between
the NLIE excited states and the scattering theory of an underlying
field theory. The integral terms in the NLIE and in the energy formula
go as $O(e^{-l})$ and can be discarded. In the case of a single hole
placed at $\vartheta_{1}$, the NLIE becomes\[
Z(\vartheta_{1})=2l\sinh\vartheta_{1}+\chi(2\vartheta_{1})+P\left(\vartheta_{1}|H_{+},H_{-}\right)=2l\sinh\vartheta_{1}+\mathcal{F}(\vartheta_{1},H_{+})+\mathcal{F}(\vartheta_{1},H_{-})\]
with\[
\mathcal{F}(\vartheta,H)=2\pi\int_{0}^{\vartheta}dx\left[F(x,H)+\int_{-\infty}^{+\infty}\frac{dk}{2\pi}e^{ikx}\frac{\sinh\frac{3\pi}{2}k\sinh\frac{\pi}{2}(p-1)k}{\sinh\frac{\pi}{4}pk\sinh\pi k}\right]\]
Interpreted, along lines of analysis very similar to those of \cite{FRT2,feverati},
as a quantization rule for momentum of a particle of energy $\mathcal{M}\cosh\vartheta_{1}$,
this yields its reflection amplitudes at both boundaries, given by
$\mathcal{R}(\vartheta,\xi_{\pm})=e^{i\mathcal{F}(\vartheta,H_{\pm})}$.
Known identities \cite{LSSM} allow to identify such reflection matrix
with the Ghoshal Zamolodchikov soliton-soliton one, upon suitable
identification of the parameters $H_{\pm}$ and $\xi_{\pm}$. $\xi_{\pm}$
in turn are connected to the DSG boundary parameters $\phi_{\pm}$
\cite{ghoshal-zam} thus allowing us to relate $H_{\pm}$ and $\phi_{\pm}$
as\[
H_{\pm}=p(1\mp\frac{8}{\beta}\phi_{\pm})\quad.\]
Notice that the periodicity $\phi_{\pm}\to\phi_{\pm}+\frac{2\pi}{\beta}$
reflects in the periodicity $H_{\pm}\to H_{\pm}\pm2(p+1)$. By changing
sign to both magnetic fields $h_{\pm}\to-h_{\pm}$ simultaneously,
one can check that also the antisoliton reflection matrix is correctly
reproduced. 

We also checked, along similar lines, that by considering suitable
combinations of holes and non-magnetic complex roots we reproduce
the correct bulk S-matrix as well as the expected Ghoshal-Zamolodchikov
reflection matrices in some simple multiparticle states.

Our analysis is limited to the case of boundary magnetic fields $|h_{\pm}|<h_{c}=1+\cos\gamma$
so that there is no way to accommodate any magnetic root. This corresponds
to absence of boundary bound states, as observed in \cite{saleur-skorik}.
All bulk states then scatter through the unexcited Ghoshal Zamolodchikov
reflection matrix. To go beyond this $h_{c}$ limitation, one should
modify the Fourier representation of $F$, i.e. consider analytic
continuation. The hope is that such analytic continuation should naturally
introduce source terms for magnetic roots in $g(\vartheta|\{\vartheta_{k}\})$
in accordance to Saleur Skorik analysis of Bethe equations, where
they found that vacuum changes by adding the maximal string of magnetic
roots. Boundary bound states should then be obtained by removing some
of these magnetic roots, thus creating holes on the imaginary axis.
This procedure turns out quite cumbersome and delicate, and needs
more investigation, so we choose to postpone the treatment of boundary
bound states to a future more detailed paper \cite{preparation}.
In the IR limit, however, by dropping the convolution term in NLIE,
it is easy to check that one can reproduce the Mattson Dorey excited
reflection matrices.

\section{UV limit and conformal theory}

In the limit $l\rightarrow0$ we make contact with the UV regime of
DSG. The roots and holes may rescale to infinity or stay in a finite
region as\[
\vartheta=\vartheta^{\pm}\pm\log\frac{1}{l}\quad\mathrm{or\quad}\vartheta=\vartheta^{0}\quad.\]
Therefore, we classify these into three types which will be denoted
by indices {}``$\pm,\,0$''. Accordingly, the NLIE splits into left
and right kink equations. Standard manipulations \cite{ddv97,feverati}
lead to the following energy formula\[
E(L)\sim\frac{\pi}{L}\left(\Delta-\frac{1}{24}\right)\]
with\begin{equation}
\Delta=\frac{p}{p+1}\left[\frac{1}{2}\left(\frac{H_{+}+H_{-}}{2p}-1-2(S-S^{+})-2(K-K_{W}^{+})\right)+\frac{p+1}{2p}S\right]^{2}+N\label{eq:delta}\end{equation}
and

\[
N=\left(I_{H}^{+}-I_{C}^{+}-I_{W}^{+}-2I_{S}^{+}+L_{W}^{+}S^{+}+2K-2K_{W}^{+}-S^{+}-2\left(S^{+}\right)^{2}\right)\in\mathbb{Z}\]
where $I_{A}^{+}=\sum_{k}I_{A,k}^{+}$ with $A=H,C,W$... the various
types of roots and holes, and \[
2S^{+}=N_{H}^{+}-2N_{S}^{+}-M_{C}^{+}-2M_{W}^{+}\mathrm{step}(p-1)\quad,\quad L_{W}^{+}=\mathrm{sign}(p-1)(M_{W}-M_{W}^{+})\quad.\]
The integers $K$ and $K_{W}^{+}$ are defined through the equations\[
Z_{+}(-\infty)=2\mathrm{Im}\log[1-(-1)^{M_{SC}}e^{iZ_{+}(-\infty)}]+\pi+2\pi K\]
\[
g_{+}(-\infty|\{\vartheta_{k}^{+}\})=2\chi(\infty)\left(S-2S^{+}\right)+2\pi K_{W}^{+}\quad.\]

This result should be compared with $c=1$ CFT of a boson with Dirichlet
conformal boundary conditions which is compactified on a circle of
radius $R=\frac{\sqrt{4\pi}}{\beta}$ \cite{affleck-oshikawa-saleur,saleur-review}.
The Hilbert space is composed of Heisenberg algebra representations
$\mathcal{Q}_{m}$ of $c=1$ CFT whose primary states $|m\rangle$
are created from the vacuum by vertex operators $:e^{i\kappa\phi}:$
of $U(1)$ charge \[
\kappa=\frac{1}{2}\left(\frac{\phi_{+}-\phi_{-}}{\sqrt{\pi}}+\frac{1}{2}mR\right)\]
and conformal dimensions $\Delta_{m}=2\kappa^{2}$. All other states
are created from these ones by applying repeatedly the creation operators\[
\mathcal{Q}_{m}=\{ a_{-k_{1}}...a_{-k_{p}}|m\rangle,\, k_{1},...,k_{p}\in\mathbb{Z}_{+}\}\quad.\]
 For a generic state $|i\rangle\in\mathcal{Q}_{m}$ the energy is
given by\[
E_{i}=\frac{\pi}{L}\left(\Delta_{m}+N_{i}-\frac{1}{24}\right)\quad,\quad N_{i}=\sum_{j=1}^{p}k_{p}\in\mathbb{Z}_{+}\]
Comparing this formula with eq.(\ref{eq:delta}) we see that the winding
number $m$ is related to the XXZ spin by $m=2S$. Notice that, according
to this formula, the ground state ($m=0$) with Dirichlet boundary
condition has not conformal dimension 0 as in the periodic CFT: the
nontrivial boundaries contribute some energy to the Casimir effect.

We end this section by presenting, only in graphical form (fig. 1),
a simple example of numerical integration of NLIE for the vacuum and
few solitonic excited states. A detailed numerical comparison of these
and other numerical data with a Truncated Conformal Space Approach
is planned in \cite{preparation}. 

\begin{center}\includegraphics{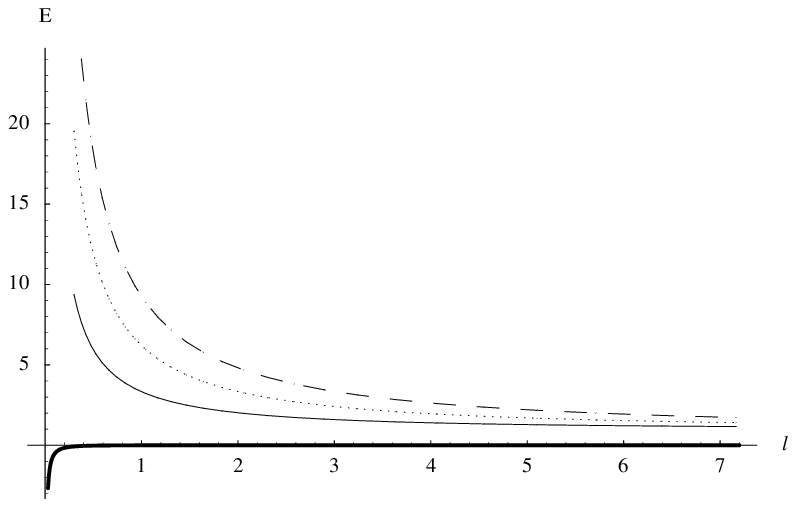}\end{center}

{\small Figure 1: Casimir energy levels for the vacuum state (thick
line) and soliton states with quantum number $I=1$ (filled line),
$I=2$ (dotted line), $I=3$ (dotted-dashed line). In this example
$p=2.3$, $H_{+}=2.2$ and $H_{-}=2.4$}{\small \par}

\section{Conclusions and perspectives}

In this letter we have obtained the NLIE governing the finite size
effects for excited states in sine-Gordon field theory with two boundaries,
each with an independent Dirichlet boundary condition as a continuum
limit of the Bethe ansatz equations of the alternating inhomogeneous
XXZ spin chain. Analysis of the IR and UV limits gives strong evidence
that the underlying model is actually the DSG theory. Understanding
of states describing the scattering of bulk solitonic particles with
Ghoshal Zamolodchikov unexcited boundaries are under control.

A better understanding of the generation mechanism of boundary bound
states for $|h_{\pm}|>h_{c}$ as analytic continuation of the NLIE
due to the singularities of the boundary source terms $F$, and of
the general structure of vacuum in this case, could lead to a full
control of the excited boundary situations too, which represents an
achievement of crucial importance in the framework of NLIE approach.

Also, other bulk states should be analyzed more carefully, as breather
scattering off unexcited or excited boundaries. In this preliminary
study we did not perform numerical analysis of NLIE. It should however
be very valuable in itself and even better if comparable with suitable
truncated conformal space approach data.

Finally the understanding of this Dirichlet boundary conditions case
should be seen just as a step towards the full investigation of NLIE
for general integrable boundary conditions in sine-Gordon model, whose
route has been recently opened by the deduction of vacuum low magnetic
field NLIE in \cite{ahn-nepomechie} starting from the Bethe ansatz
for non diagonal boundary conditions proposed and studied in \cite{nepomechie1,nepomechie2,nepomechie-ravanini,cao-lin-shi-wang}.

\section*{Acknowledgments}

We would like to dedicate this paper to the memory of Prof. Sung-Kil
Yang. In particular FR has been, years ago, collaborator of Sung-Kil
and remembers with the highest respect his great human qualities and
the exemplar professionnality of a first class scientist that we all
miss.

We acknowledge the participants of the APCTP Focus Program 2003 for
the interest in this work and valuable discussions. FR thanks the
APCTP and Ewha Womans University, MB the Department of Mathematics
at University of York and CA the INFN section of Bologna for kind
hospitality during this work. 

This work is supported by the EU research network EUCLID (Grant no.
HPRN-CT-2002-00325) and the INFN Iniziativa Specifica TO12 (M.B. and
F.R.) and by Korean Research Foundation 2002-070-C00025 (C.A.).

\end{document}